# Shear-wave manipulation by embedded soft devices

Linli Chen, Chao Ma, Pingping Zheng, Qian Zhao, Zheng Chang[*]

College of Science, China Agricultural University, Beijing 100083, China

**Abstract**

Hyperelastic transformation theory has proven shear-wave manipulation devices with various functions can be designed by utilizing neo-Hookean material with appropriate pre-deformation. However, it is still elusive that how can such devices match with the background medium in which they embedded. In this work, we present a systematic formulation of the transmission and reflection of elastic waves at the interface between un-deformed and pre-deformed hyperelastic materials. With the combination of theoretical analyses and numerical simulations, we specifically investigate the shear-wave propagation from an un-deformed neo-Hookean material to the one subject to different homogeneous deformations. Among three typical deformation modes, we found "constrained" uniaxial tension and simple shear guarantee total transmission, whereas "ordinary" uniaxial tension and hydrostatic compression cause wave reflection. On this basis, three embedded shear-wave manipulation devices, including a unidirectional cloak, a splicable beam bend, and a concave lens, are proposed and verified through numerical simulations. This work may pave the way for the design and realization of soft-matter-based wave control devices. Potential applications can be anticipated in nondestructive testing, structure impact protection, biomedical imaging, and soft robotics.

[*] Author to whom correspondence should be addressed. Electronic mail: changzh@cau.edu.cn (Z. Chang).

# 1. Introduction

Elastic waves are mechanical vibrations that propagate in solid media. As a common process of energy and information transmission, they have been intensively studied over centuries [1, 2] and found technologically significant applications in many branches of engineering, such as nondestructive testing, medical imaging, and geophysical prospecting. In recent years, hyperelastic soft materials, such as elastomers and gels, have drawn much attention in elastodynamics [3-9], for their high sensitivity, diverse material behavior, and reversible geometry effects accompany finite deformation. By virtue of these features, many soft elastic-wave-control devices with tunable or adaptive properties, e.g. invisibility cloak [10-12], wave mode splitter [13], and phononic crystal [14, 15], have been proposed. Compared to "hard" devices, such soft devices have a natural advantage of integration with other soft-systems, and may yield new insight into the designs of biomedical and soft robotic technologies.

Usually, a soft device doesn't work independently, while is embedded in its "working environment" or the background medium. Although previous studies on soft devices have focused on exploring their wave control capacities, little research has been devoted to matching characteristics with the periphery. This is a key issue that has been overlooked because it directly affects the quality factor of the soft devices.

The small-on-large theory [16] provides a natural framework to analyze the problem of incremental linear wave motions superimposed onto a finite pre-deformation. This supplies an important basis not only for the soft device design, but also for the investigation of their matching properties. However, in the theory, pre-deformed hyperelastic material usually possesses an effective anisotropy and the behavior of a *Cosserat*-like continuum [17]. Under this circumstance, the classical theory [2] of the transmission and reflection of elastic waves becomes incompetent. Little is known regarding the principle that dictates such a physical process.

Recently, based on the small-on-large theory, a hyperelastic transformation theory (HTT) [4] has been proposed. It reveals the hyperelastic soft materials with specified strain energy functions (SEF) can behave like smart transformation metamaterials [18] and realize some unique wave manipulation properties by tuning their deformation. In particular, it shows that in neo-Hookean materials, shear-wave (S-wave) paths deform in accordance with the distorted material curve [13]. This finding enables us to design S-wave control devices by introducing pre-deformation in neo-Hookean materials. Compare to the traditional transformation technique [19-21], HTT eliminates the requirement of microstructures. Therefore, the corresponding soft devices exhibit remarkable potential for non-dispersion and broadband wave manipulation. For such "transformation devices", the matching between the transformation domain and the background medium is also of great significance. Open questions include how to find appropriate deformations by which the deformed neo-Hookean material is matched with a second medium, and how to design soft devices that simultaneously acquire wave control function and matching property.

To address the aforementioned issues, in this work, we present a systematic formulation of the transmission and reflection of elastic waves at the interface between un-deformed and pre-deformed hyperelastic materials. Specifically, we investigate S-wave propagate through an interface between un-deformed neo-Hookean material and the one subject to several typical deformations (uniaxial tension, hydrostatic compression, and simple shear). On this basis, we propose three embedded soft devices, including a unidirectional invisibility cloak, a splicable beam bend, and a concave lens. We also perform both theoretical analyses and numerical simulations to demonstrate the efficiency of such devices.

The paper is arranged as follows. In Section 2, we briefly review the small-on-large theory as preliminary. In Section 3, we investigate the transmission and reflection of elastic waves at the interface between un-deformed and pre-deformed hyperelastic materials. Particularly, we explore the transmission characteristics of the S-wave at the interface of the neo-Hookean materials. In Section 4, we propose three soft devices for S-wave manipulation. Finally, we close with our brief concluding remarks and a discussion on the avenues for future work in Section 5.

## 2. Small-on-large theory: linear elastic wave propagation in a finitely deformed hyperelastic material

For a hyperelastic solid with the strain energy function $W$, the equilibrium equation of the finite deformation can be written as

$$(C_{ijkl}U_{l,k})_{,i} = 0, \tag{1}$$

where $\mathbf{U}$ denotes the finite displacement, $C_{ijkl} = \partial^2 W / \partial F_{ji} \partial F_{lk}$ is the component of the fourth-order elastic tensor $\mathbf{C}$ expressed in the initial configuration, and $F_{ij} = \partial \tilde{x}_i / \partial \tilde{X}_j$ the deformation gradient. $\tilde{X}_j$ and $\tilde{x}_i$ are the material coordinates in the un-deformed and the deformed configurations, respectively. Further, the incremental wave motion $\mathbf{u}$ superimposed onto the finite deformation $\mathbf{U}$ is governed by

$$(C_{0i'jk'l}u_{l,k'})_{,i'} = \rho_0 \ddot{u}_j, \tag{2}$$

in time domain, or

$$(C_{0i'jk'l}u_{l,k'})_{,i'} = -\omega^2 \rho_0 u_j, \tag{3}$$

in frequency domain, with a pushing forward operation on the elastic tensor $\mathbf{C}$ and the initial mass density $\rho$, i.e. [16]

$$C_{0i'jk'l} = J^{-1} F_{i'i} F_{k'k} C_{ijkl}, \quad \rho_0 = J^{-1}\rho, \tag{4}$$

where $J = \det(\mathbf{F})$ is the volumetric ratio.

For a homogenously deformed hyperelastic material, incremental plane waves in time domain can be expressed in the form of

$$u_i = A_i e^{\mathrm{i}(kl_j \cdot x_j - \omega t)}, \tag{5}$$

in which $\mathbf{A}$ is the wave amplitude, $\mathrm{i}$ denotes the imaginary unit, $k$ is the wave number, $\mathbf{l}$ is the unit vector in the wave direction and $\omega$ is the angular frequency. By inserting Eq.(5) into Eq.(2), the *Christoffel* equation can be obtained as [2]

$$C_{0i'jk'l} l_{i'} l_{k'} m_l = c^2 \rho_0 m_j, \tag{6}$$

where $\mathbf{m}$ is a unit polarization vector, $c = \omega/k$ denotes the phase velocity of the elastic wave. By solving the eigenvalue problem of Eq.(6), we can obtain the phase velocities ($V_\mathrm{P}$ and $V_\mathrm{S}$) and the polarization directions ($x'_i$) of the longitudinal (P-) and shear (S-) waves, all of which are necessary in the following derivation.

## 3. Transmission and reflection of elastic waves at the interface between un-deformed and pre-deformed hyperelastic materials

*3.1 Theoretical formulae*

We consider a two-dimensional (in-plane) problem of a plane elastic wave incident on a plane interface between un-deformed (Domain I, $\mathbf{F} = $ unit tensor) and pre-deformed (Domain II, $\mathbf{F} \neq $ unit tensor) hyperelastic materials, as illustrated in Fig.1. For simplicity, normal incidence is considered to avoid mode conversion [2] induced by oblique incidence. Meanwhile, the deformation in Domain II is considered to be homogeneous, so that the elastic waves travel in straight paths.

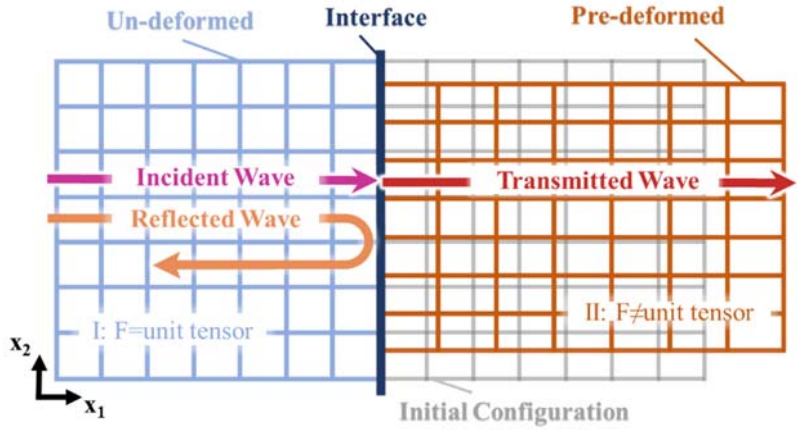

Fig.1. Schematic diagram of an elastic wave propagations from an un-deformed hyperelastic material (I, light blue mesh) to a pre-deformed one (II, orange mesh). At the interface (dark blue line), only normal incidence is considered. The initial configuration (gray mesh) of the pre-deformed material is also presented as a reference.

At the interface, the particle velocity $v_i = \partial u_i / \partial t$ and the stress $\sigma_{ij} = C_{0ijkl} u_{l,k}$, that induced from the elastic wave motion $\mathbf{u}$ should be continuous, i.e.

$$v_i^{\mathrm{I}} = v_i^{\mathrm{II}}, \qquad (7)$$

$$\sigma_{ij}^{\mathrm{I}} = \sigma_{ij}^{\mathrm{II}}. \qquad (8)$$

For an S-wave propagates alone the $x_1$-direction (see Fig.1) and polarizes in $x_2$-direction, according to Eqs.(5) and (6), the particle velocity and the stress at the

interface ($x_1 = a$) in Domain I can be expressed as

$$\left(v_2^{\mathrm{I}}\right)_{inc} = -\mathrm{i}\omega A e^{\mathrm{i}(kl_1 \cdot a - \omega t)}, \quad \left(\sigma_{12}^{\mathrm{I}}\right)_{inc} = -\mathrm{i}\omega A \rho V_S e^{\mathrm{i}(kl_1 \cdot a - \omega t)}, \tag{9}$$

in which the subscript $inc$ indicates the incident wave, the subscript $i = 1, 2$ denotes the component of the spatial coordinate $\mathbf{x}$, and $A$ is the scalar wave amplitude. Similarly, the particle velocity and the stress of the P- and S-waves emerged during reflection and transmission at the interface can be expressed as

$$\left(v_1^{\mathrm{I}}\right)_R = -\mathrm{i}\omega B e^{\mathrm{i}(kl_1 \cdot a - \omega t)}, \quad \left(\sigma_1^{\mathrm{I}}\right)_R = -\mathrm{i}\omega B \rho V_P^{\mathrm{I}} e^{\mathrm{i}(kl_1 \cdot a - \omega t)}, \tag{10}$$

$$\left(v_2^{\mathrm{I}}\right)_R = -\mathrm{i}\omega C e^{\mathrm{i}(kl_1 \cdot a - \omega t)}, \quad \left(\sigma_{12}^{\mathrm{I}}\right)_R = -\mathrm{i}\omega C \rho V_S^{\mathrm{I}} e^{\mathrm{i}(kl_1 \cdot a - \omega t)}, \tag{11}$$

$$\left(v_{1'}^{\mathrm{II}}\right)_T = -\mathrm{i}\omega D e^{\mathrm{i}(kl_1 \cdot a - \omega t)}, \quad \left(\sigma_{1'}^{\mathrm{II}}\right)_T = \mathrm{i}\omega D \rho_0 V_P^{\mathrm{II}} e^{\mathrm{i}(kl_1 \cdot a - \omega t)}, \tag{12}$$

$$\left(v_{2'}^{\mathrm{II}}\right)_T = -\mathrm{i}\omega E e^{\mathrm{i}(kl_1 \cdot a - \omega t)}, \quad \left(\sigma_{1'2'}^{\mathrm{II}}\right)_T = \mathrm{i}\omega E \rho_0 V_S^{\mathrm{II}} e^{\mathrm{i}(kl_1 \cdot a - \omega t)}, \tag{13}$$

where the subscripts R and T indicate the reflection and the transmitted waves, the subscript $i' = 1', 2'$ denotes the polarization coordinate $\mathbf{x}'$, while $B$, $C$, $D$ and $E$ are the scalar amplitudes of the waves.

Due to the equivalent anisotropy induced from the finite deformation, quasi-mode elastic waves may exist in pre-deformed hyperelastic materials [22]. Therefore, we introduce a polarization angle $\varphi$ as the angle between the polarization direction $x'_i$ and the spatial direction $x_i$. In this fashion, Eq.(12) and Eq.(13) can be expressed in terms of the spatial coordinate $\mathbf{x}$. By inserting Eqs.(10)-(13) into Eq.(7), we can obtain

$$B = D\cos\varphi - E\sin\varphi, \tag{14}$$

$$A + C = D\sin\varphi + E\cos\varphi. \tag{15}$$

Similarly, Eq.(8) can be expressed as

$$B\rho V_P^{\mathrm{I}} = -D\rho_0 V_P^{\mathrm{II}} \cos\varphi + E\rho_0 V_S^{\mathrm{II}} \sin\varphi, \tag{16}$$

$$-A\rho V_S^{\mathrm{I}} + C\rho V_S^{\mathrm{I}} = -D\rho_0 V_P^{\mathrm{II}} \sin\varphi - E\rho_0 V_S^{\mathrm{II}} \cos\varphi. \tag{17}$$

By solving Eqs.(14)-(17), the transmission and reflection coefficients $\Gamma_{ns}^m$, can be obtained as

$$\Gamma_{TP}^{S} = \frac{D}{A} = \frac{2\rho V_S^I \sin\varphi (\rho_0 V_S^{II} + \rho V_P^I)}{\Delta}, \tag{18}$$

$$\Gamma_{TS}^{S} = \frac{E}{A} = \frac{2\rho V_S^I \cos\varphi (\rho V_P^I + \rho_0 V_P^{II})}{\Delta}, \tag{19}$$

$$\Gamma_{RP}^{S} = \frac{B}{A} = -\frac{2\rho\rho_0 V_S^I \sin\varphi \cos\varphi (V_P^{II} - V_S^{II})}{\Delta}, \tag{20}$$

$$\Gamma_{RS}^{S} = \frac{C}{A} = \frac{\rho^2 V_S^I V_P^I - \rho_0^2 V_S^{II} V_P^{II} + \rho\rho_0 \sin^2\varphi (V_S^I V_S^{II} - V_P^I V_P^{II}) + \rho\rho_0 \cos^2\varphi (V_S^I V_P^{II} - V_S^{II} V_P^I)}{\Delta},$$

$$\tag{21}$$

where $\Delta = \rho^2 V_S^I V_P^I + \rho_0^2 V_S^{II} V_P^{II} + \rho\rho_0 \sin^2\varphi (V_S^I V_S^{II} + V_P^I V_P^{II}) + \rho\rho_0 \cos^2\varphi (V_S^I V_P^{II} + V_S^{II} V_P^I)$.

For $\Gamma_{ns}^{m}$, the superscript $m = P, S$ denotes the incident wave mode, the subscript $n = T, R$ denotes the transmitted and reflection waves, while the subscript $s = P, S$ denotes the wave modes.

Similarly, for P-wave incidence, the transmission and reflection coefficients can be obtained as

$$\Gamma_{TP}^{P} = \frac{2\rho V_P^I \cos\varphi (\rho_0 V_S^{II} + \rho V_S^I)}{\Delta}, \tag{22}$$

$$\Gamma_{TS}^{P} = -\frac{2\rho V_P^I \sin\varphi (\rho V_S^I + \rho_0 V_P^{II})}{\Delta}, \tag{23}$$

$$\Gamma_{RP}^{P} = \frac{\rho^2 V_S^I V_P^I - \rho_0^2 V_S^{II} V_P^{II} + \rho\rho_0 \sin^2\varphi (V_P^I V_P^{II} - V_S^I V_S^{II}) + \rho\rho_0 \cos^2\varphi (V_S^{II} V_P^I - V_S^I V_P^{II})}{\Delta},$$

$$\tag{24}$$

$$\Gamma_{RS}^{P} = -\frac{2\rho\rho_0 V_P^I \sin\varphi \cos\varphi (V_P^{II} - V_S^{II})}{\Delta}. \tag{25}$$

*3.1.1 A particular case: S-wave incidence in a neo-Hookean material*

As an example of the aforementioned theory, meanwhile, as the theoretical basis for designing a neo-Hookean transformation device, we consider in the following the transmission and reflection of an S-wave at the interface between an un-deformed neo-

Hookean material and a pre-deformed one. The two-dimensional SEF of the neo-Hookean material can be written as [23]

$$W = \frac{\lambda}{2}(J-1)^2 - \mu \ln(J) + \frac{\mu}{2}(I_1 - 2), \tag{26}$$

where $I_1$ is the first invariant of the right Cauchy-Green tensor, $\lambda$ and $\mu$ are the *Lamé* constants.

It is noticeable that the P- and S-waves propagate in a pre-deformed neo-Hookean material in their pure modes [22]. Whereupon, Eqs.(1)-(4) can be simplified as

$$\Gamma_{RP}^{S} = \frac{B}{A} = 0, \tag{27}$$

$$\Gamma_{RS}^{S} = \frac{C}{A} = \frac{\rho V_S^{I} - \rho_0 V_S^{II}}{\rho V_S^{I} + \rho_0 V_S^{II}}, \tag{28}$$

$$\Gamma_{TP}^{S} = \frac{D}{A} = 0, \tag{29}$$

$$\Gamma_{TS}^{S} = \frac{E}{A} = \frac{2\rho V_S^{I}}{\rho V_P^{I} + \rho_0 V_P^{II}}. \tag{30}$$

Eq.(27) and Eq.(29) denote that no P-wave generated during the S-wave transmission. From Eq.(28) and Eq.(30), one can find $\Gamma_{TS}^{S} - \Gamma_{RS}^{S} = 1$ and $\Gamma_{RS}^{S} \leq 0$. The negative reflection coefficient is owing to the half-wave loss, indicating the wave receives a 180° phase shift.

*3.2 Numerical method*

To validate the aforementioned theoretical results and the performances of the subsequent soft devices, numerical simulations have been performed by a two-step model using the software COMSOL Multiphysics.

In the first step, the finite deformation of a hyperelastic material, which is governed by Eq.(1), is calculated with the structural mechanics module. Consequently, the deformed geometry, together with the deformation gradient **F**, is imported into the wave field analysis.

In the second step, i.e. the wave field analysis, both steady-state and transient-state analyses are utilized. Usually, the steady-state analysis intuitively indicates the wave field distribution and the direction of wave propagation, while the transient-state analysis unambiguously distinguishes the incident and reflected waves.

In the steady-state analysis, Eq.(3) is modeled with the weak form PDE interface to deal with the asymmetry of the elastic tensor $\mathbf{C}_0$ (see Eq.(4)). A portion of pre-deformed neo-Hookean material or a designed soft device (with $\mathbf{F}$ obtained from the first step) is embedded in an un-deformed neo-Hookean domain, as an example demonstrated in Fig.2(b). On the periphery of the un-deformed domain, perfectly matched layers (not shown) [24] are applied to avoid unnecessary reflection. An S-wave Gaussian beam is imported at an appropriate location as needed.

In the transient-state analysis, weak form PDE is also employed to solve Eq.(2). As shown in Fig.2(c), two rectangular domains are applied as the un-deformed and pre-deformed neo-Hookean domains. A harmonic plane S-wave with the excitation lasts for three wavelengths, are imported at the left boundary. The upper and lower boundaries are set to be *Floquet* periodic, and the right boundary is left to be free.

In the following simulations, we choose $\lambda = 4.32$ MPa, $\mu = 1.08$ MPa, and $\rho = 1050 \text{ kg/m}^3$ as the initial material parameters of the neo-Hookean material, which refer to a compressible variant of material PSM-4 [3]. The amplitude of the wave source is set as $A = 0.01$ m. The angular frequencies are set as $\omega_{st} = 3$ kHz and $\omega_{tr} = 0.4$ kHz in steady-state and transient-state analyses, respectively.

### *3.3 Particular cases for the pre-deformation*

To validate the theoretical results and obtain some prior knowledge for the soft device design, three typical modes of pre-deformation, including uniaxial tension, hydrostatic compression, and simple shear, have been considered.

### *3.3.1 Uniaxial tension*

In the first case, we consider the hyperelastic materials subject to uniaxial tension. The deformation can be accomplished by applying a displacement $U_1 = 2.667$ m on the right boundary of a $1 \text{ m} \times 1 \text{ m}$ material domain, with the rest boundaries set to be rollers, as shown in Fig.2(a). The deformation gradient of such "constrained" uniaxial tension can be described as $F_{11} = 1.667$, $F_{22} = 1$ and $F_{12} = F_{21} = 0$. Correspondingly, the elongation ratio in $x_1$-direction is $\eta = 1.667$.

For an S-wave horizontally propagates through the pre-deformed neo-Hookean material, the steady-state displacement field $|u_2|$ has been illustrated in Fig.2(b). It shows the S-wave is not altered by the two interfaces it passes through, which means the "constrained" uniaxial tensioned neo-Hookean material perfectly matches with the un-deformed one. Such numerical result coincides with the theoretical calculation obtained from Eq.(28) and Eq.(30), i.e. $\Gamma_{TS}^{S}=1$ and $\Gamma_{RS}^{S}=0$. In transient-state analysis, two typical snapshots ($t=0.01\,\text{s}$ and $t=0.029\,\text{s}$) of the normalized displacement field $D_2 = u_2/A$ are displayed in Fig.2(c). In the two snapshots, the distributions of $D_2$ along the $x_1$-direction are illustrated in Fig.2(d), with the theoretical results marked as horizontal dashed lines. It indicates the wavelength in the deformed domain at $t=0.029\,\text{s}$, is $\eta=1.667$ times of that in the un-deformed one at $t=0.01\,\text{s}$. After the wave impinging at the interface, there is no backward wave generates.

If we relax the upper and lower roller constrains in the "constrained" uniaxial tension, the uniaxial tension becomes an "ordinary" one, as shown in Fig.2(e). Correspondingly, the deformation gradient turns to be $F_{11}=1.667$, $F_{22}=0.685$, and $F_{12}=F_{21}=0$. In this case, the perfect matching ceases to exist. In the steady-state wave field demonstrated in Fig.2(f), it can be observed that the field strength in the pre-deformed neo-Hookean material is higher than that in the un-deformed domain, due to the superposition of incident and reflected waves. The mismatch can be clearly illustrated in the transient-state analysis (Figs.2(g) and (h)). At $t=0.029\,\text{s}$, the amplitude of the transmitted wave is significantly lower than that of the incident one, and the reflection can be clearly observed in the un-deformed domain. Both the steady-state and transient-state analyses confirm the theoretical results $\Gamma_{TS}^{S}=0.813$ and $\Gamma_{RS}^{S}=-0.187$.

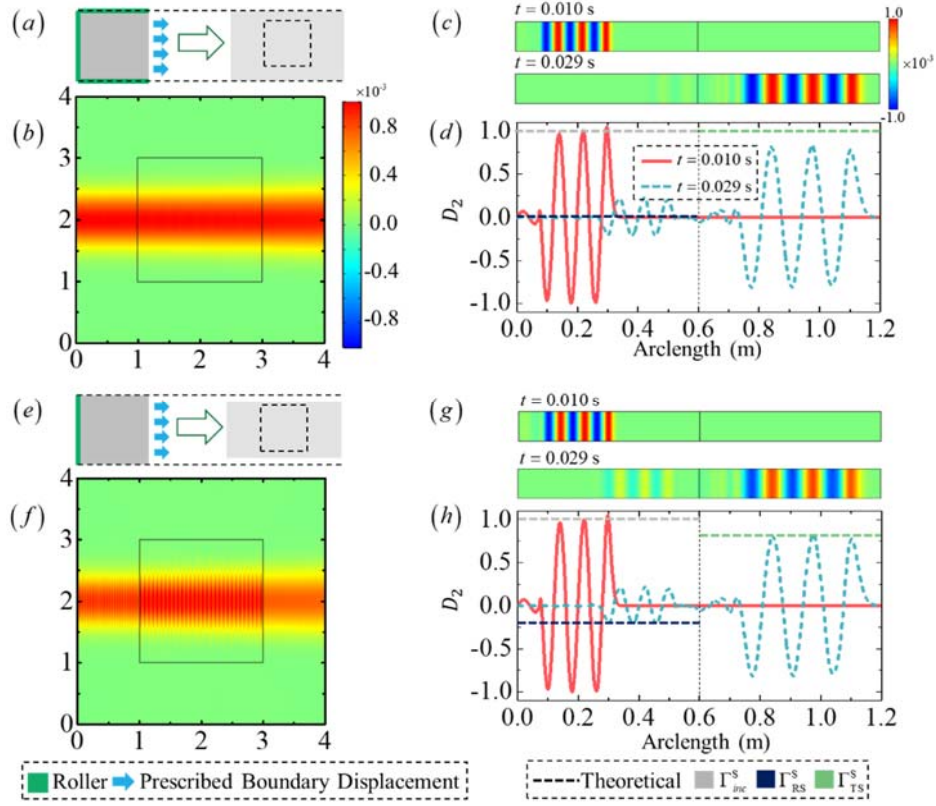

Fig.2. Transmission and reflection of elastic waves at the interfaces between undeformed and uniaxial-tensioned hyperelastic materials. (a)-(d), "constrained" uniaxial-tension; (e)-(h), "ordinary" uniaxial-tension. (a) and (e) are the schematic diagrams of the deformation modes. (b) and (f) are the steady-state displacement fields of $|u_2|$, when the shear-wave beams are incident on the square pre-deformed domains. (c) and (g) are the transient-state displacement fields of $D_2$ at $t = 0.01\,\text{s}$ and $t = 0.029\,\text{s}$. (d) and (h) depict the amplitudes of $D_2$ at the two snapshots, together with the theoretical results of $\Gamma_{int}^S$, $\Gamma_{RS}^S$, $\Gamma_{TS}^S$.

### 3.3.2 Hydrostatic compression

In the second case, we consider the hydrostatic compression. For a square neo-Hookean domain with a side length of $1\,\text{m}$, the deformation can be accomplished with the left and lower boundaries set to be rollers, and simultaneously, the upper and right boundaries set as the prescribed displacements of $U_1 = U_2 = -0.5\,\text{m}$, as shown in Fig.3(a). In this fashion, the deformation gradient is $F_{11} = F_{22} = 0.875$ and

$F_{12} = F_{21} = 0$. Although it can hardly be preserved in the steady-state wave field (Fig.3(b)), the transient result (Fig.3(c) and (d)) manifests a slight impedance mismatch, as predicted by the theoretical results $\Gamma_{TS}^{S}=0.933$ and $\Gamma_{RS}^{S}=-0.067$.

*3.3.3 Simple shear*

In the third case, simple shear deformation is considered. As demonstrated in Fig.3(e), the deformation can be implemented by applying the prescribed body displacement $U_2 = x/3$ m on the square material domain. The corresponding deformation gradient is $F_{11} = F_{22} = 1$, $F_{21} = 0.333$ and $F_{12} = 0$. As depicted in Fig.3(f), the wave beam has been shifted together with the simple shear deformation, demonstrating the S-wave manipulation capability of the neo-Hookean material. Meanwhile, no reflection occurs when the wave beam propagates through the material, as shown in both Figs.3(f)-(h).

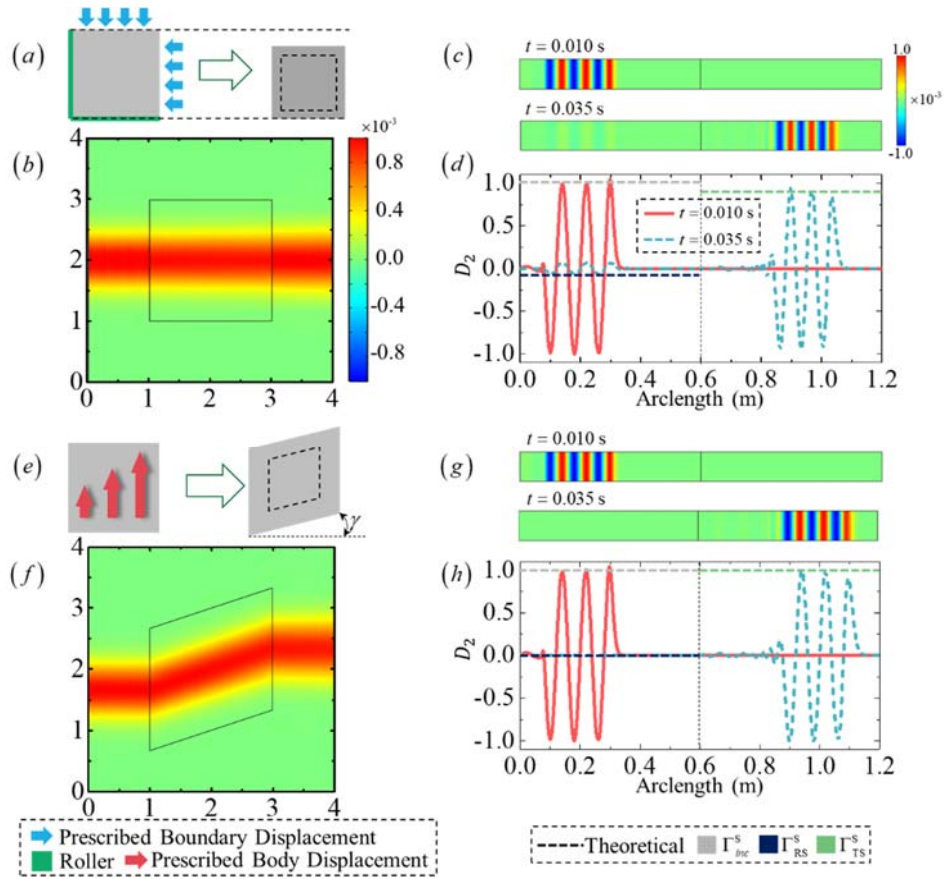

Fig.3. Transmission and reflection of elastic waves at the interfaces between undeformed hyperelastic material and the one subject to (a)-(d) hydrostatic compression

and (e)-(h) simple shear. (a) and (e) are the implementation schemes of the deformation modes. (b) and (f) are the steady-state displacement fields of $|u_2|$, when the shear-wave beams are incident on the square pre-deformed domains. (c) and (g) are the transient-state displacement fields of $D_2$ at $t = 0.01\,\text{s}$ and $t = 0.035\,\text{s}$. (d) and (h) depict the amplitudes of $D_2$ at the two snapshots, together with the theoretical results of $\Gamma_{int}^S$, $\Gamma_{RS}^S$, and $\Gamma_{TS}^S$.

## 4. Embedded neo-Hookean transformation devices for S-wave manipulation

Aforementioned theoretical and numerical investigations provide a guide for the design of embedded soft devices. In the following, three devices are proposed as examples.

*4.1 Unidirectional cloak*

Consider a circular cavity with radius $r = 0.05\,\text{m}$ in an un-deformed neo-Hookean domain. It causes strong scattering when an S-wave beam with the corresponding wavelength of $l = 0.1\,\text{m}$ passes through, as shown in Fig.4(a). To suppress the scattering, we simply embed a neo-Hookean material which subject to "constrained" uniaxial tension with the enlarge ratio $\eta$ into a rectangular domain around the cavity, as shown in Figs.4(b) and (c).

The distribution of the displacement field $u_2$ for the case of $\eta=2$ is plotted in Fig.4(b). It is shown the reflectionless of the cloak on its left and right interfaces. Meanwhile, the wavelength becomes $\eta$ times as long as the original one in the cloak region, and the $u_2$ field out of the cloak region looks smoother than that in the case without the cloak. It is predictable the larger the elongation ratio $\eta$, as the characteristic radius $r/(\eta l)$ of the cavity decreases, the more effectively scattering is suppressed. If we construct the cloak with "constrained" uniaxial tensioned neo-Hookean material with the elongation ratio of $\eta=10$, as demonstrated in Fig.4(c), the scattering is difficult to perceive and the wave field out of the cloak region looks similar to that in the free space without cavity, as shown in Fig.4(d).

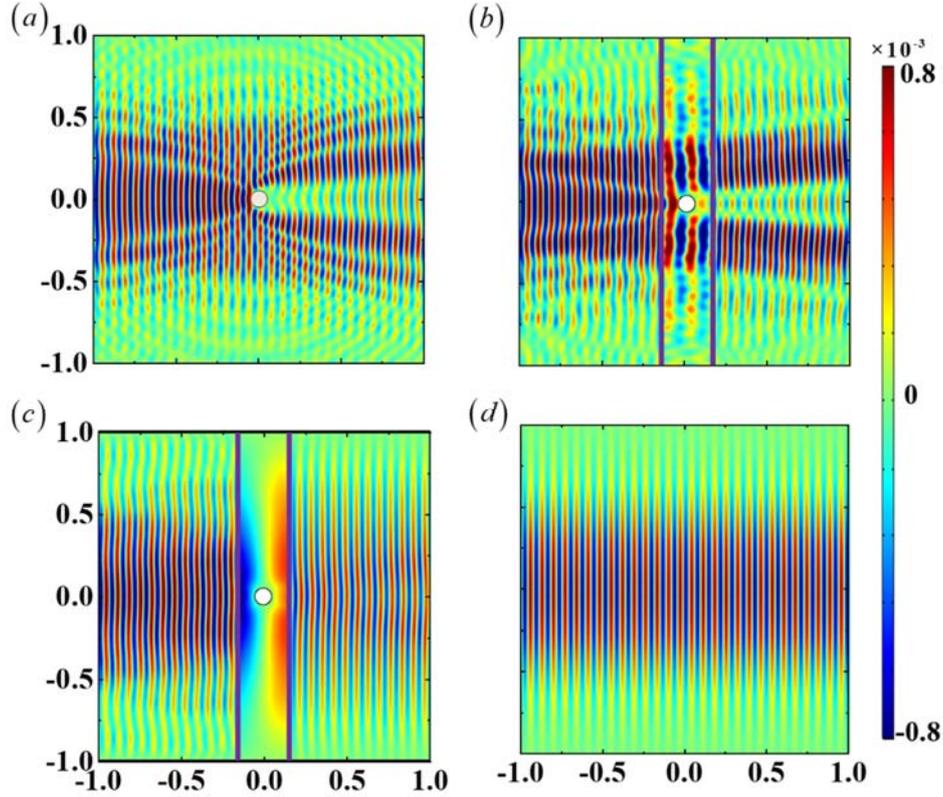

Fig.4. Displacement field $u_2$ in neo-Hookean material with and without the unidirectional cloak for the incidence of an S-wave beam. (a) uncloaked case: a cavity without cloak. (b) cloaked case: a cavity with a cloak of $\eta=2$. (c) cloaked case: a cavity with a cloak of $\eta=10$. (d) reference case: free space.

*4.2 Splicable beam bend*

By invoking HTM, an S-wave beam bend can be constructed by bending a rectangular neo-Hookean material to a certain angle. However, when the angle is large and the aspect ratio of the rectangle is relatively small, instability and damage may happen in the soft material. To avoid such failure, we propose a splicable beam bend, by which a large bending angle can be achieved through the assembling of several moderately deformed components. A schematic diagram of a splicable $\pi/2$-bend is shown in Fig.5(a). The beam bend consists of two identical trapezoid components. Each of them is achieved by fixing one of the long sides of a rectangular neo-Hookean material with length $l=1.5\,\text{m}$ and width $d=1\,\text{m}$ and applying an appropriate boundary displacement (for example, $U_2 = -(x_1 - 1.5)$ for the component I, see Fig.5(a) ) to the opposite side, meanwhile constraining the two short sides with rollers.

The curl field of an S-wave beam propagate through the beam bend are demonstrated in Fig.5(b), indicating an ideal wave manipulation. At the seam of the two components of the beam bend, material properties are identical, thus no reflection occurs. At the inlet and outlet interfaces, the deformation is inhomogeneous. To render the matching properties, the transmission and reflection coefficients of six infinitesimal micro-elements on the inlet ($L_1$ in Fig.5(b)) are analyzed, as shown in Fig.5(c). Both theoretical results and numerical simulations show the transmission coefficient is universally close to 1, while the reflection coefficient is negligible. The transient analyses of the S-wave behavior on the micro-element at $(0.6, 2.75)$, which is the midpoint of $L_1$, are indicated in Fig.5(d). At $t = 0.0284 \text{ s}$, the wave propagate through the interface with the wavelength of 1.81 times as long as the original one. Meanwhile, no obvious reflection can be perceived. By polar decomposing the deformation gradient ($F_{11} = 1.81$, $F_{12} = -2.01 \times 10^{-6}$, $F_{21} = 0.336$ and $F_{22} = 1$), we find the micro-element is a "constrained" uniaxial-tensioned one ($\eta = 1.81$), with a negligible rotation angle of $2.1 \times 10^{-4} \pi$.

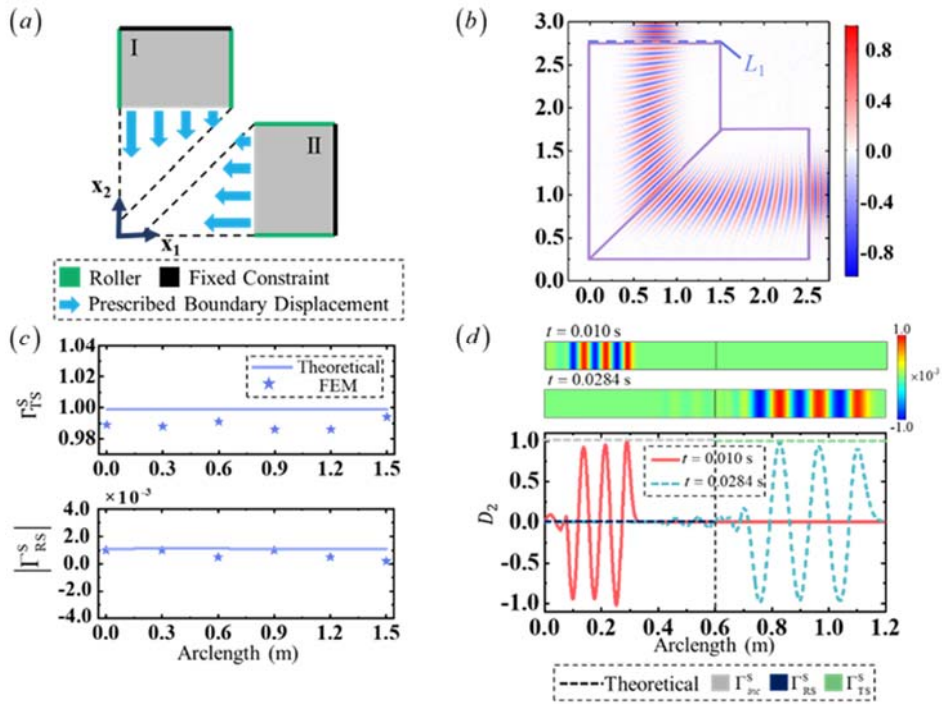

Fig.5. Schematic diagram and performance of the splicable beam bend. (a) The implementation scheme of the beam bend. (b) The curl field in and out of the beam bend when an S-wave beam incident in $-x_2$-direction. (c) Numerical simulation and theoretical prediction of the transmission $\left(\Gamma_{TS}^S\right)$ and reflection $\left(\left|\Gamma_{RS}^S\right|\right)$ coefficients of the micro-elements, on auxiliary segment $L_1$ shown in (b). (d) Transient-state displacement fields of $D_2$ at $t = 0.01\,\text{s}$ and $t = 0.0284\,\text{s}$, together with the theoretical results of $\Gamma_{int}^S$, $\Gamma_{RS}^S$, and $\Gamma_{TS}^S$, at the point of $(0.6, 2.75)$ (midpoint of $L_1$) in (b).

*4.3 Concave lens*

In the last example, we propose a concave lens, which can focus a plane S-wave to a point, and thus covert a plane wave into a cylindrical one. Consider a rectangular neo-Hookean material with length $l = 3\,\text{m}$ and width $d = 6\,\text{m}$. As depicted in Fig.6(a), we fix the left boundary, and constrain the upper and lower boundaries with rollers. To focus the wave at $(6, 0)\,\text{m}$ from the lens, as shown in Fig.6(b), we apply a displacement of $U_1 = 3 - \sqrt{18 - x_2^2}$ at the right boundary of the neo-Hookean domain.

As the curl field plotted in Fig.6(b), the horizontally propagated plane wave beam is focused at the expected position, and then radiates as a cylindrical wave. The transmission and reflection coefficients on micro-elements of the two interfaces (see $L_1$ and $L_2$ in Fig.6(b)) where the wave beam passes through have been examined. With a similar deformation mode as the aforementioned beam bend, the micro-elements on $L_1$ indicate an approximate match with the background medium, as shown in Fig.6(c). However, along the arclength of $L_2$, mismatch appears at positions away from the axis of symmetry, as shown in Fig.6(d). $8\%$ of the transmission loss can be perceived at $1.6\,\text{m}$ away from the axis of symmetry. Nevertheless, the performance of the lens can be guaranteed with the satisfactory transmission coefficient in the region where the beam energy is mainly distributed.

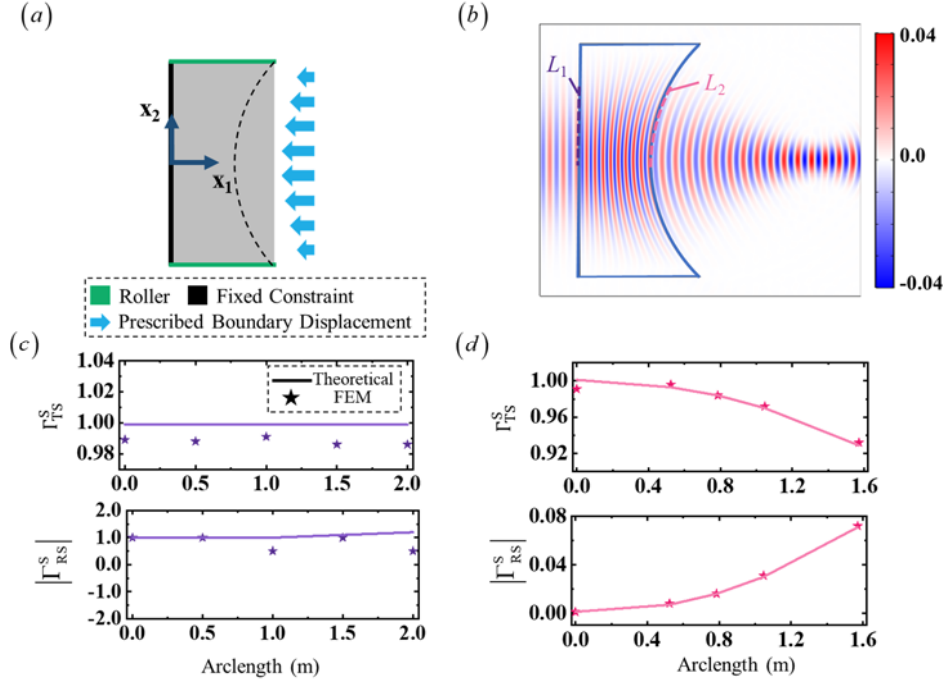

Fig.6. Schematic diagram and performance of the concave lens. (a) The implementation scheme of the lens. (b) The curl field in and out of the lens when an S-wave beam incident in $x_1$-direction. (c) and (d) are the numerical simulation and theoretical prediction of the transmission $\left(\Gamma_{TS}^S\right)$ and reflection $\left(\left|\Gamma_{RS}^S\right|\right)$ coefficients of the micro elements on auxiliary segments $L_1$ and $L_2$, respectively.

## 5. Conclusion and Discussion

In this work, we provide a systematic formulation of the transmission and reflection of elastic waves at an interface between un-deformed and pre-deformed hyperelastic materials. In addition to such material, it is worth noting that the formulation can also be applied to other media with *Cosserat* form. For the concise of the paper, we only consider the normal incidence of elastic waves. More sophisticated but fruitful results can be expected from a systematic investigation of oblique incidence.

It has shown the neo-Hookean material with the SEF of Eq.(26) exhibits an outstanding character in the design of soft devices, as in which the elastic waves propagate in pure-modes. On this basis, to make the S-wave device match with the background medium, we can either utilize the "constrained" uniaxial tension or simple shear deformation at the interface (perfect match), or keep the input and output interface un-deform (perfect match) or deform as small as possible (approximate match). Although all the three devices proposed in this paper are designed basing on the "constrained" uniaxial tension, the case of simple shear deformation can be referenced in a design of wave-mode splitter which has been reported in a previous contribution. In the next step, further effort will be directed towards the realization of the mechanical loading proposed in this work. It is also a challenging issue as it requires the loading devices do not affect wave propagation.

We hope this work may provide some new insight into energy and information transmission in such fields as in non-destructive testing, structure impact protection, biomedical imaging or soft robotics.

## Acknowledgement

This work was supported by the National Natural Science Foundation of China (Grant No. 11602294) and the Chinese Universities Scientific Fund (Grant No. 2019TC134).

# References

1. Achenbach J 2012 *Wave propagation in elastic solids* (Amsterdam: Elsevier)

2. Auld B A 1974 *Acoustic Fields and Waves in Solids* (USA: Wiley)

3. Bertoldi K and Boyce M C 2008 Wave propagation and instabilities in monolithic and periodically structured elastomeric materials undergoing large deformations *Phys. Rev.* B **78** 184107

4. Norris A N and W J Parnell Hyperelastic cloaking theory: transformation elasticity with pre-stressed solids *Proc. R. Soc.* A **468** 2881

5. Guo D K, Chen Y, Chang Z and Hu G K 2017 Longitudinal elastic wave control by pre-deforming semi-linear materials *J. Acoust. Soc. Am* **142** 1229

6. Xin F X and Lu T J 2017 Self-controlled wave propagation in hyperelastic media *Sci. Rep.* **7** 7581

7. Li G Y, He Q, Mangan R, Xu G Q, Mo C, Luo J W, Destrade M and Cao Y P 2017 Guided waves in pre-stressed hyperelastic plates and tubes: Application to the ultrasound elastography of thin-walled soft materials *J. Mech. Phys. Solids* **102** 67

8. Zhang K, Ma C, He Q, Lin S T, Chen Y, Zhang Y, Fang N X and Zhao X H 2019 Metagel with Broadband Tunable Acoustic Properties Over Air–Water–Solid Ranges *Adv. Funct. Mater.* 1903699.

9. Deng B, Zhang Y, He Q, Tournat V, Wang P and Bertoldi K 2019 Propagation of elastic solitons in chains of pre-deformed beams *New J. Phys.* **21** 073008

10. Parnell W J, Norris A N and Shearer T 2012 Employing pre-stress to generate finite cloaks for antiplane elastic waves *App. Phys. Lett.* **100** 171907

11. Zhang P and Parnell W J 2018 Hyperelastic antiplane ground cloaking *J. Acoust. Soc. Am.* **143** 2878

12. Guo D K, Chang Z and Hu G K 2019 In-Plane Semi-Linear Cloaks with Arbitrary Shape. *Acta Mech. Solida Sin.* **32** 277

13. Chang Z, Guo H Y, Li B and Feng X Q 2015 Disentangling longitudinal and shear elastic waves by neo-Hookean soft devices *App. Phys. Lett.* **106** 161903

14. Liu Y, Chang Z and Feng X Q 2016 Stable elastic wave band-gaps of phononic crystals with hyperelastic transformation materials *Extr. Mech. Lett.* **11** 37

15. Zhang P and Parnell W J 2017 Soft phononic crystals with deformation-independent band gaps *Proc. R. Soc.*A **473** 20160865

16. Ogden R W 2007 Incremental Statics and Dynamics of Pre-Stressed Elastic Materials *Waves in Nonlinear Pre-Stressed Materials* ed Destrade M and Saccomandi G (Vienna: Springer) p 1

17. Cosserat E and Cosserat F 1909 *Théorie des corps déformables* (Hermann)

18. Shin D, Urzhumov Y, Jun Y, Kang G, Baek S, Choi M, Park H, Kim K and Smith D R 2012 Broadband electromagnetic cloaking with smart metamaterials *Nat. Comm.* **3** 1213

19. Pendry J B, Schurig D and Smith D R 2006 Controlling electromagnetic fields *Science* **312** 1780

20. Rahm M, Cummer S A, Schurig D, Pendry J B and Smith D R 2008 Optical design of reflectionless complex media by finite embedded coordinate transformations *Phys. Rev. Lett.* **100** 063903

21. Norris A N and Shuvalov A L 2011 Elastic cloaking theory *Wave Motion* **48** 525

22. Chen L, Chang Z and Qin T 2017 Elastic wave propagation in simple-sheared hyperelastic materials with different constitutive models *Int. J. Solids and Struct.* **126** 1

23. Ogden R W and Sternberg E 1985 *Nonlinear Elastic Deformations* (New York Dover)

24. Chang Z, Guo D K, Feng X Q and Hu G K 2014 A facile method to realize perfectly matched layers for elastic waves *Wave Motion* **51** 1170